# Electronic rotons and Wigner crystallites in a two-dimensional dipole liquid

Soobin Park[1], Minjae Huh[1], Chris Jozwiak[2], Eli Rotenberg[2], Aaron Bostwick[2] & Keun Su Kim[1]*

[1]Department of Physics, College of Science, Yonsei University, Seoul, Korea.
[2]Advanced Light Source, E. O. Lawrence Berkeley National Laboratory, Berkeley, CA, USA.
*e-mail: keunsukim@yonsei.ac.kr

A key concept proposed by Landau to explain superfluid liquid helium is the elementary excitation of quantum particles called rotons[1-8]. The irregular arrangement of atoms in a liquid forms the aperiodic dispersion of rotons that played a pivotal role in understanding fractional quantum Hall liquid (magneto-rotons)[9,10] and the supersolidity of Bose-Einstein condensates[11-13]. Even for a two-dimensional electron or dipole liquid in the absence of a magnetic field, their repulsive interactions were predicted to form a roton minimum[14-19] that can be used to trace the transition to Wigner crystals[20-24] and superconductivity[25-27], but it has not been observed. Here, we report the observation of such electronic rotons in a two-dimensional dipole liquid of alkali-metal ions doping charges to surface layers of black phosphorus. Our data reveal a striking aperiodic dispersion of rotons characterised by a local minimum of energy at a finite momentum. As the density of dipoles decreases, where interactions dominate over kinetic energy, the roton gap reduces to 0 as in crystals, signalling Wigner crystallisation. Our model shows the importance of short-range order arising from repulsion between dipoles, which can be viewed as the formation of Wigner crystallites (bubbles or stripes) floating in the sea of Fermi liquids. Our results reveal that the primary origin of electronic rotons (and the pseudogap) is strong correlations.

Quantum states of matter arising from the correlation of many particles have been one of the grand challenges in physics. Motivated by the discovery of liquid $^4$He with no viscosity, Landau proposed a theory of superfluidity based on the concept of rotons[1,2]. As shown in Fig. 1a, the energy of excitations in $^4$He first increases with momentum ($k$), passes through a local maximum (maxon), and decreases to a local minimum (roton). The quantum theory of rotons was developed by Feynman[3] suggesting that it is observable by inelastic neutron scattering. The roton minimum was indeed observed in liquid $^4$He (neutral boson)[4] and $^3$He (neutral fermion)[5], but the microscopic origin of rotons has been a long-standing puzzle[6]: The first interpretation by Landau, Onsager, and Feynman is associated with the rotational motion of atoms in the liquid state (the so-called ghost of a vanished vortex ring)[1-3,6], which is why it was named 'roton'. On the other hand, this idea was challenged by another point of view that the roton minimum is a precursor of crystallisation[7,8] (the so-called ghost of a Bragg spot), which may also be related to a local charge-density-wave instability[7].

This concept of rotons also played a pivotal role in understanding the fractional quantum Hall effect, another quantum state of matter arising from the interplay of two-dimensional (2D) electrons in the presence of a strong perpendicular magnetic field[9,10]. The theory of collective excitations in fractional quantum Hall states was proposed to exhibit the similar roton minimum termed magneto-rotons[9], which has been observed in experiments[10] and known to arise from density waves. The deepening of this roton minimum was interpreted as a barometer of its stability to trigger the transition from quantum Hall liquids to Wigner crystals[9], but has not been directly observed in experiments. Recently, this roton instability was invoked in Bose-Einstein condensates (alkali metals) to search for supersolidity[11-13].

Even for the 2D quantum fluid of electrons (charged fermions) or dipoles (charged bosons) in the absence of a magnetic field, their repulsive interaction was predicted to form a roton minimum not only in collective excitations[14-19], but also in the single-particle spectrum[28-31]. This electronic analogue of rotons with no magnetic field, termed 'electronic rotons' here, is important to explore a phase transition from Fermi liquids to novel quantum states, such as Wigner crystals[20-24], superfluidity (superconductivity)[25,26], and supersolidity[27]. Along the boundary of these phases, it has been predicted that there is no first order phase transition, and there should exist intermediate microemulsion phases (or electronic liquid crystals) in the form of bubbles or stripes whose size can be comparable to the interparticle spacing[23]. However, it has been challenging to observe electronic rotons due to the lack of a suitable experimental system combined with a high-resolution spectroscopy.

Our experimental system is the 2D dipole liquid or the electron-hole bilayer of a low-density alkali metal (Na, K, Rb, and Cs) placed on a layered crystalline insulator (black phosphorus), as schematically illustrated in Fig. 1b. The previous results[32-34] have shown that alkali metals are energetically in favour of staying on the surface rather than penetrating into the bulk. Each alkali metal donates a charge to black phosphorus due to their difference in electron affinity, being ionised to form the heavy hole layer. To screen these positively charged ions, the doped electrons as the only free charge carriers are concentrated close to the surface of black phosphorus, as manifested by a reduction in the band gap[32,33]. This in turn forms a mass-anisotropic 2D dipole layer or an electron-hole bilayer with the vertical distance of $d$ (Fig. 1b), which can also be viewed as a 2D electron gas under the influence of scattering by the potential of alkali-metal ions (it is more so at the higher density[33]).

The in-plane diffusion barrier of alkali metals on black phosphorus with their density in the order of $10^{13}$ cm$^{-2}$ is known to be extremely weak. Indeed, as shown by scanning tunnelling microscopy[34], each dipole migrates even at the temperature of 15 K (hence, dipole liquid). There is no long-range order of dipoles, but their repulsive interactions render the system to have the average distance between the dipoles as $a$, which is what we mean by short-range order. This is reflected in the structure factor $S(k)$ by a broad peak at $k = 2\pi/a$, and a set of these peaks forms a ring (Fig. 1b, inset) that is anisotropic due to the armchair-zigzag crystal structure of black phosphorus[34]. The presence of this broad peak in $S(k)$ was found responsible for back-bending dispersion and pseudogap[33]. This is therefore an ideal system, combined with angle-resolved photoemission spectroscopy (ARPES), to explore the phase diagram of 2D dipoles as a function of $d/a$ and $a/a_B^*$ ($r_s$)[22-27] where $a_B^*$ is the effective Bohr radius, as summarised in Fig. 1c.

## Observation of electronic rotons

ARPES data of black phosphorus doped by K with the 2D dipole density $n = 3.8 \times 10^{13}$ cm$^{-2}$ is shown in Fig. 1d along the zigzag ($y$) direction. As expected for doped black phosphorus, there is a well-defined quadratic dispersion of the conduction band with the effective mass $m^* = 1.25 m_e$, where $m_e$ is the electron rest mass. Near the Fermi energy ($E_F$), a pseudogap $\Delta_{PG}$ arising from the localisation of electrons by the scattering potential of K ions manifests with the magnitude of 77 meV in agreement with that reported before[33]. To see how this pseudogapped band dispersion is connected away from $\pm k_{PG}$, we normalised the intensity of energy-distribution curves (EDCs) at each $k_y$ point (Extended Data Fig. 1 for raw data). As shown in Fig. 1e, the normalised data of Fig. 1d reveals a striking aperiodicity in the band dispersion: The energy and $k$ distances between local minima (0, 2, and 4) and between local maxima (1 and 3) are not equal to each other. It is consistently found in ARPES data taken in armchair ($x$) and diagonal ($s$) axes (Extended Data Fig. 2), and for those doped by Na, Rb, and Cs (Extended Data Fig. 3). This aperiodic dispersion of electrons is impossible for crystalline solids, and even in liquid metals[35], quasicrystals[36], and amorphous solids[37], no such aperiodicity has been observed.

The curve fitting of EDCs with the Gaussian function more clearly shows the relative energy of their peaks as a function of $k_y$, as shown in Fig. 2a. It first increases with $k_y$, pass through a local maximum at $k_y = 0.21$ Å$^{-1}$, and decreases to a local minimum at $k_y = 0.42$ Å$^{-1}$, which is the characteristic dispersion similar to Landau's dispersion of rotons (Fig. 1a)[28,29]. By taking the energy of this roton minimum relative to that at $k_y = 0$ Å$^{-1}$, we estimate the roton gap ($\Delta_R$) to be about 30 meV at $3.8 \times 10^{13}$ cm$^{-2}$. This roton minimum was proposed by Feynman[3] to be intimately associated with the average distance ($a$) between particles in a liquid state (Fig. 2b), which is represented by a broad peak in $S(k)$ at $k = 2\pi/a$. In this theory[3], the single-particle energy ($\hbar^2 k^2 / 2m^*$) renormalised by $1/S(k)$ equals to the energy spectrum $E(k)$ as

$$E(k) = \frac{\hbar^2 k^2}{2m^* S(k)}.$$

Using this relation and $E(k)$ obtained by ARPES in Fig. 2a, $S(k)$ is experimentally determined as shown in Fig. 2c. There is a broad peak whose height is $I_R$, which confirms the existence of short-range order with $a = 9.4$ Å. With this $a$ divided by $a_B^*$, where $a_B^* = 4\pi \hbar^2 \varepsilon / m^* e^2$ and $\varepsilon$ is the dielectric constant, we estimate the dimensionless coupling parameter $r_s$ to be 7.2.

To model the observed spectrum, we employed an analytic theory developed for the band structure of non-crystalline systems in the presence of short-range order[30,31]. This model is concerned with the effect of scattering by Dirac's $\delta$-function or a spherical potential well whose width is $W$ and depth is $D$. The spatial distribution of these scattering potentials can be approximated by the correlation function $g(r)$ with only the first order peak at $a$ that is rectangular in shape with the width of $b$ as shown by inset in Fig. 2b. The broadness of this peak in $g(r)$ is tuned by the dimensionless parameter $p = b/a$. The presence of this peak in $g(r)$, which is related to $S(k)$ by Fourier transformation, is responsible for renormalisations of a single-particle dispersion[28,29] similar to Feynman's theory[3]. A series of $E(k)$'s calculated by raising $p$ is shown in Fig. 2d (Methods), where we find the clear dependence of $\varDelta_R$ on $p$, as indicated by the arrow in inset and plotted in Fig. 2e as a function of $p$. This was used to fit $\varDelta_R$ in our data (Fig. 2a), which nicely reproduces the observed band dispersion as shown by the overlaid black line (Extended Data Fig. 4 for ARPES simulations).

**Observation of Wigner crystallisation**

An interesting possibility arising from the discovery of electronic rotons is to explore their instability against the formation of Wigner crystals by raising $r_s$ in the experimental system, as shown by the phase diagram in Fig. 1c. In our experiments, $r_s$ can be tuned based on the relation $\pi r_s^2 a_B^{*2} = 1/n$ by varying the density of alkali metals or $n$. The three representative ARPES data taken at different $n$ are displayed in Fig. 3a-c. We found that $\varDelta_R$ at $n = 5.2 \times 10^{13}$ cm$^{-2}$ (Fig. 3a) is decreased at $n = 3.8 \times 10^{13}$ cm$^{-2}$ (Fig. 3b) and becomes even a negative value close to zero at $n = 2.1 \times 10^{13}$ cm$^{-2}$ (Fig. 3c). This is a striking tendency that the kinetic energy of electrons ($E_{KE}$) is dominated by the potential energy of their repulsive interactions ($E_{PE}$), which is a clear signature of the transition to Wigner crystals, that is, Wigner crystallisation.

For a quantitative analysis, we employed curve fitting to a more complete doping series of ARPES data (Extended Data Fig. 5) with $E(k)$ calculated by the analytic model[30,31] explained above (Methods). The variation of $\varDelta_R$ and $r_s$ obtained from this curve fitting is plotted as a function of $n$ in Fig. 3d,e. We estimate $n$ at which $\varDelta_R$ reverses its sign to be $2.4 \times 10^{13}$ cm$^{-2}$, as indicated by the dotted line in Fig. 3d. As also marked by the same dotted line in Fig. 3e, this $n$ was used to estimate the critical $r_s$ of Wigner crystallisation, denoted as $r_{sc}$, to be 9.2. This is much smaller than $r_{sc} = 37$ predicted for the pure 2D electron liquid ($d/a > 1$, Fig. 1c)[21] and $r_{sc} = 20$ predicted for the pure 2D dipole liquid ($d/a < 1$, Fig. 1c)[22], but comparable to the $r_{sc}$ of 11 observed for Wigner crystallisation in transition-metal dichalcogenides[38] and that of 4.8–10 observed in the 2D electron gas of Si[39], GaAs[40], and ZnO[41]. This can be naturally explained by the effect of natural defects or impurities[42], which is estimated by theoretical calculations for our system to reduce $r_{sc}$ to the range of 6.7–11.8 (Methods).

It would be interesting to trace how the microscopic structure of 2D dipole liquids changes in the process of Wigner crystallisation. Figure 3f shows a doping series of $S(k)$ taken from ARPES data with positive $\Delta_R$ through Feynman's relation. We used curve fitting to each $S(k)$ with the Gaussian function and backgrounds, and the variation of $I_R$ is plotted as a function of $n$ in Fig. 3g. The peak in $S(k)$ located near $k = 2\pi/a$ develops as indicated by the red arrow, when the system approaches Wigner crystallisation. A consistent picture can be identified separately in the doping dependence of $p$ in $g(r)$. Based on the relation between $p$ and $\Delta_R$ in Fig. 2e, the variation of $p$ obtained from Fig. 3d is plotted as a function of $n$ in Fig. 3h. We found the narrowing of the peak in $g(r)$ at $r = a$, signalling that the distance between two neighbouring dipoles becomes more uniform at incipient Wigner crystallisation (red circle).

Our results provide a key clue to the microscopic origin of rotons that has been debated in the study of superfluidity[6-8]. It is important to note that the spectrum of electronic rotons can be reproduced by the analytic model, which is essentially based on the first peak in $g(r)$ with no need to include the higher order. It suggests the formation of 'Wigner crystallites', as schematically illustrated in Fig. 2b, where a cluster of atoms forms a local microcrystal[17] with a well-defined interatomic distance of $a$ within the area enclosed by the light red circle. Outside of this circle, the arrangement of atoms is loosely correlated to those in the circle, which is averaged out to be a uniform background in $g(r)$. This is a quantum state of matter with only short-range order or with broken local translational symmetry, but the randomly oriented Wigner crystallites make the system have no global rotational symmetry breaking.

The Wigner crystallite can be viewed as a seed of crystals floating in the sea of Fermi liquids, which grows in size as the system undergoes a phase transition to Wigner crystals. This in fact reminds of the theoretical prediction for 2D dipole systems that there is no first order phase transition between Fermi liquids and Wigner crystals[23]. Rather, there should be a variety of intermediate phases or electronic liquid crystals termed 'microemulsion phases' as a mixture of two competing phases in the form of bubbles or stripes. Especially if $d$ is comparable to $a$ as in our case, there is no stable Wigner crystals with long-range order by quantum fluctuations (quantum melting)[23]. There remain bubbles or stripes of the Winger crystals whose size is of the order of interparticle spacing ($a$), which is naturally connected to the concept of Wigner crystallites in the sea of Fermi liquids (Fig. 2b).

**Correlations as the origin of rotons**

It is important to clarify the primary origin of short-range order in Wigner crystallites. If the constituent particles were arbitrarily distributed without any interactions between them, which is the case of a gas, no peak is expected for $g(r)$ and $S(k)$. However, if the long-range Coulomb ($1/r$) or quasi-long-range dipole-dipole ($1/r^3$) repulsive interactions came into play, there must be a well-defined length scale of $a$, within which the particles cannot be placed as in the hard sphere model. This is exactly what the broad peak in $g(r)$ at $a$ (Fig. 3h) or that in $S(k)$ at $2\pi/a$ (Fig. 3g) represents[16-18]. That is, the formation of short-range order in Wigner crystallites cannot be explained in the single-particle picture, but is primarily due to strong correlations (repulsive interactions) between dipoles at the range of $n$ studied here.

In fact, this short-range order of dipoles with the average distance of $a$ is the key factor in understanding the pseudogap observed at $E_F$ and $k = \pi/a$[33]. The size of $\Delta_{PG}$ in Fig. 1d is found more or less the same even in the process of Wigner crystallisation (Extended Data Fig. 5). If the energy cost to break local translational symmetry could be compensated by opening pseudogap at $E_F$ as in glassy charge density waves, the formation of short-range order may also be a result of competing long-range repulsive and short-range attractive interactions, which has been predicted to yield bubble or clump phases[43-45] similar to Wigner crystallites. In any cases, the primary origin for electronic rotons as well as the pseudogap is essentially strong correlations between the constituent quantum particles.

Lastly, it is important to mention how the aperiodic band dispersion revealed here (Fig. 1e) is related to the back-bending dispersion and the pseudogap[33]. In the electronic structure of liquid metals, there are multiple branches of one band in terms of partial wave analysis[46], as summarised in Extended Data Fig. 6. The *p*-wave or *d*-wave state at resonance scattering leads to a branch with back-bending dispersion and pseudogap, which was first introduced by Anderson and McMillan[47]. However, the model improved by Schwartz and Ehrenreich[48] showed that there is another branch that extends towards the zone boundary. This in fact corresponds to *s*-wave states[49], for which resonance scattering is forbidden, and the effect of multiple scattering leads to the aperiodic (damped oscillatory) dispersion[30,31] exactly as observed in this work. Therefore, our results provide the complete picture for the spectral function of non-crystalline systems in the presence of short-range order.

The observation of electronic rotons opens a few exciting opportunities: It is important to explore the two-roton bound state[50] that has been proposed as a key concept in the study of superfluidity and factional quantum Hall effect. As $d/a$ in our system is tuneable by $n$ as well as different kinds of elements, it would also be interesting to see if this system can be driven into the excitonic regime ($d/a < 1$, Fig. 1c), where the exotic ground states associated to Bose-Einstein condensation were predicted in theory[25-27]. This is therefore a simple and easily tuneable platform to examine the theory of strongly interacting quantum particles.

**References**

1. Landau, L. Theory of the Superfluidity of Helium II. *Phys. Rev.* **60,** 356–358 (1941). https://doi.org/10.1103/PhysRev.60.356
2. Landau, L. On the theory of superfluidity of helium II. *J. Phys. U.S.S.R.* **11,** 91 (1947). https://doi.org/10.1016/B978-0-08-010586-4.50068-7
3. Feynman, R. P. & Cohen, M. Energy Spectrum of the Excitations in Liquid Helium. *Phys. Rev.* **102,** 1189–1204 (1956). https://doi.org/10.1103/PhysRev.102.1189
4. Henshaw, D.G. & Woods, A. D. B. Modes of Atomic Motions in Liquid Helium by Inelastic Scattering of Neutrons. *Phys. Rev.* **121,** 1266–1274 (1961). https://doi.org/10.1103/PhysRev.121.1266
5. Godfrin, H. *et al*. Observation of a roton collective mode in a two-dimensional Fermi liquid. *Nature* **483,** 576–579 (2012). https://doi.org/10.1038/nature10919
6. Donnelly, R. Rotons: a low-temperature puzzle. *Phys. World* **10,** 25–30 (1997). https://doi.org/10.1088/2058-7058/10/2/24
7. Nozières. P. Is the Roton in Superfluid $^4$He the Ghost of a Bragg Spot? *J. Low Temp. Phys.* **137,** 45–67 (2004). https://doi.org/10.1023/B:JOLT.0000044234.82957.2f
8. Bobrov, V., Trigger, S. & Litinski, D. Universality of the Phonon-Roton Spectrum in Liquids and Superfluidity of $^4$He. *Z. Naturforsch.* **71,** 565–575 (2016). https://doi.org/10.1515/zna-2015-0397
9. Girvin, S. M., MacDonald, A. H. & Platzman, P. M. Magneto-roton theory of collective excitations in the fractional quantum Hall effect. *Phys. Rev. B* **33,** 2481–2494 (1986). https://doi.org/10.1103/PhysRevB.33.2481
10. Kukushkin, I. V., Smet, J. H., Scarola, V. W., Umansky, V. & von Klitzing, K. Dispersion of the Excitations of Fractional Quantum Hall States. *Science* **324,** 1044–1047 (2009). https://www.science.org/doi/10.1126/science.1171472
11. Mottl, R. *et al*. Roton-Type Mode Softening in a Quantum Gas with Cavity-Mediated Long-Range Interactions. *Science* **336,** 1570–1573 (2012). https://doi.org/10.1126/science.1220314
12. Chomaz, L. *et al.* Observation of roton mode population in a dipolar quantum gas. *Nat. Phys.* **14,** 442–446 (2018). https://doi.org/10.1038/s41567-018-0054-7
13. Mukherjee, B. *et al.* Crystallization of bosonic quantum Hall states in a rotating quantum gas. *Nature* **601,** 58–62 (2022). https://doi.org/10.1038/s41586-021-04170-2
14. Apaja, V., Halinen, J., Halonen, V., Krotscheck, E. & Saarela, M. Charged-boson fluid in two and three dimensions. *Phys. Rev. B* **55,** 12925–12945 (1997). https://doi.org/10.1103/PhysRevB.55.12925
15. De Palo, S., Conti, S. & Moroni, S. Monte Carlo simulations of two-dimensional charged bosons. *Phys. Rev. B* **69,** 035109 (2004). https://doi.org/10.1103/PhysRevB.69.035109
16. Kalman, G. J., Hartmann, P., Golden, K. I., Filinov, A. & Donkó, Z. Correlational origin of the roton minimum. *Europhys. Lett.* **90,** 55002 (2010). https://doi.org/10.1209/0295-5075/90/55002
17. Kalman, G. J., Kyrkos, S., Golden, K. I., Hartmann, P. & Donkó, Z. The Roton Minimum: Is it a General Feature of Strongly Correlated Liquids? *Contrib. Plasma Phys.* **52,**

219–223 (2012). https://doi.org/10.1002/ctpp.201100095

18. Dorheim, T., Moldabekov, Z., Vorberger, J., Kählert, H. & Bonitz, M. Electronic pair alignment and roton feature in the warm dense electron gas. *Commun. Phys.* **5,** 304 (2022). https://doi.org/10.1038/s42005-022-01078-9

19. Lu, H., Chen, B. -B., Wu, H. -Q., Sun, K. & Meng, Z. Y. Thermodynamic Response and Neutral Excitations in Integer and Fractional Quantum Anomalous Hall States Emerging from Correlated Flat Bands. *Phys. Rev. Lett.* **132,** 236502 (2024). https://doi.org/10.1103/PhysRevLett.132.236502

20. Wigner, E. On the Interaction of Electrons in Metals. *Phys. Rev.* **46,** 1002–1010 (1934). https://doi.org/10.1103/PhysRev.46.1002

21. Tanatar, B. & Ceperley, D. M. Ground state of the two-dimensional electron gas. *Phys. Rev. B* **39,** 5005–5016 (1989). https://doi.org/10.1103/PhysRevB.39.5005

22. De Palo, S., Rapisarda, F. & Senatore, G. Excitonic Condensation in a Symmetric Electron-Hole Bilayer. *Phys. Rev. Lett.* **88,** 206401 (2002). https://doi.org/10.1103/PhysRevLett.88.206401

23. Spivak, B., Kivelson, S. A. Phase intermediate between a two-dimensional electron liquid and Wigner crystal. *Phys. Rev. B* **70,** 155114 (2004). https://doi.org/10.1103/PhysRevB.70.155114

24. Hartmann, P., Donkó, Z. & Kalman, G. J. Structure and phase diagram of strongly-coupled bipolar charged-particle bilayers. *Europhys. Lett.* **72,** 396–402 (2005). https://doi.org/10.1209/epl/i2004-10551-4

25. Lozovik, Y. E. & Yudson, V. I. A new mechanism for superconductivity: pairing between spatially separated electrons and holes. *Sov. Phys. JETP* **44,** 738–753 (1976).

26. Balatsky, A. V., Joglekar, Y. N. & Littlewood, P. B. Dipolar Superfluidity in Electron-Hole Bilayer Systems. *Phys. Rev. Lett.* **93,** 266801 (2004). https://doi.org/10.1103/PhysRevLett.93.266801

27. Joglekar, Y. N., Balatsky, A, V. & Das Sarma. S. Wigner supersolid of excitons in electron-hole bilayers. *Phys. Rev. B* **74,** 233302 (2006). https://doi.org/10.1103/PhysRevB.74.233302

28. Glyde, H. R. & Griffin, A. Zero Sound and Atomiclike Excitations: The Nature of Phonons and Rotons in Liquid $^4$He. *Phys. Rev. Lett.* **65,** 1454–1457 (1990). https://doi.org/10.1103/PhysRevLett.65.1454

29. Filinov, A. & Bonitz, M. Collective and single-particle excitations in two-dimensional dipolar Bose gases. *Phys. Rev. A* **86,** 063628 (2012). https://doi.org/10.1103/PhysRevA.86.043628

30. De Dycker, E. & Phariseau, P. On the LCAO-method for disordered materials: I. General theory. *Physica* **34,** 325–332 (1967). https://doi.org/10.1016/0031-8914(67)90046-8

31. De Dycker, E. & Phariseau, P. On the LCAO-method for disordered materials: II. Application to some simple models. *Physica* **35,** 405–416 (1967). https://doi.org/10.1016/0031-8914(67)90188-7

32. Kim, J. *et al.* Observation of tunable band gap and anisotropic Dirac semimetal state in black phosphorus. *Science* **349,** 723-726 (2015).

https://www.science.org/doi/10.1126/science.aaa6486

33. Ryu, S. H. *et al.* Pseudogap in a crystalline insulator doped by disordered metals. *Nature* **596,** 68–73 (2021). https://doi.org/10.1038/s41586-021-03683-0
34. Kiraly, B. *et al.* Anisotropic Two-Dimensional Screening at the Surface of Black Phosphorus. *Phys. Rev. Lett.* **123,** 216403 (2019). https://doi.org/10.1103/PhysRevLett.123.216403
35. Baumberger, F., Auwärter, W., Greber, T. & Osterwalder, J. Electron Coherence in a Melting Lead Mßonolayer. *Science* **306,** 2221–2224 (2004). https://doi.org/10.1126/science.1103984
36. Rotenberg, E., Theis, W., Horn, K. & Gille, P. Quasicrystalline valence bands in decagonal AlNiCo. *Nature* **406,** 602–605 (2000). https://doi.org/10.1038/35020519
37. Corbae, P. *et al.* Observation of spin-momentum locked surface states in amorphous $Bi_2Se_3$. *Nat. Mater.* **22,** 200–206 (2023). https://doi.org/10.1038/s41563-022-01458-0
38. Zhou, Y. *et al.* Bilayer Wigner crystals in a transition metal dichalcogenide heterostructure. *Nature* **595,** 48–52 (2021). https://doi.org/10.1038/s41586-021-03560-w
39. Pudalov, V. M., D'lorio, M., Kravchenko, S. V. & Campbell, J. W. Zero-magnetic-field collective insulator phase in a dilute 2D electron system. *Phys. Rev. Lett.* **70,** 1866–1869 (1993). https://doi.org/10.1103/PhysRevLett.70.1866
40. Hanein, Y., Shahar, D., Yoon, j., Li, C. C., Tsui, D. C. & Shtrikman, H. Observation of the metal-insulator transition in two-dimensional *n*-type GaAs. *Phys. Rev. B* **58,** R13338–R13340 (1993). https://doi.org/10.1103/PhysRevB.58.R13338
41. Solovyev, V. V. & Kukushkin, I. V. Renormalized Landau quasiparticle dispersion revealed by photoluminescence spectra from a two-dimensional Fermi liquid at the MgZnO/ZnO heterointerface. *Phys. Rev. B* **96,** 115131 (2017). https://doi.org/10.1103/PhysRevB.96.115131
42. Chui, S. T. & Tanatar, B. Impurity Effect on the Two-Dimensional-Electron Fluid-Solid Transition in Zero Field. *Phys. Rev. Lett.* **74,** 458–461 (1995). https://doi.org/10.1103/PhysRevLett.74.458
43. Fogler, M. M., Koulakov, A. A. & Shklovskii, B. I. Ground state of a two-dimensional electron liquid in a weak magnetic field. *Phys. Rev. B* **54,** 1853–1871 (1996). https://doi.org/10.1103/PhysRevB.54.1853
44. Reichhardt, C. J. O., Reichhardt, C. & Bishop, A. R. Structural transitions, melting, and intermediate phases for stripe- and clump-forming systems. *Phys. Rev. E* **82,** 041502 (2010). https://doi.org/10.1103/PhysRevE.82.041502
45. Pu, S., Balram, A. C., Taylor, J., Fradkin, E. & Papić, Z. Microscopic Model for Fractional Quantum Hall Nematics. *Phys. Rev. Lett.* **132,** 236503 (2024). https://link.aps.org/doi/10.1103/PhysRevLett.132.236503
46. Chang, K. S., Sher, A., Petzinger, K. G. & Weisz, G. Density of states of liquid Cu. *Phys. Rev. B* **12,** 5506–5513 (1975). https://doi.org/10.1103/PhysRevB.12.5506
47. Anderson, P. W. & McMillan, W. L. *Multiple-Scattering Theory and Resonances in Transition Metals, Proceedings of the International School of Physics "Enrico Fermi"* Course 37, edited by Marshall, W. (Academic, New York, 1967).

https://doi.org/10.1142/9789812385123_0013
48. Schwartz, L. & Ehrenreich, H. Single-Site Approximation in the Electronic Theory of Liquid Metals. *Annals of Phys.* **64,** 100–148 (1971).
https://doi.org/10.1016/0003-4916(71)90281-8
49. Morgan, G. J. Electron transport in liquid metals II. A model for the wave functions in liquid transition metals. *J. Phys. C: Solid State Phys.* **2,** 1454–1463 (1969).
https://doi.org/10.1088/0022-3719/2/8/314
50. Park, K. & Jain, J. K. Two-Roton Bound State in the Fractional Quantum Hall Effect. *Phys. Rev. Lett.* **84,** 5576–5579 (2000). https://doi.org/10.1103/PhysRevLett.84.5576

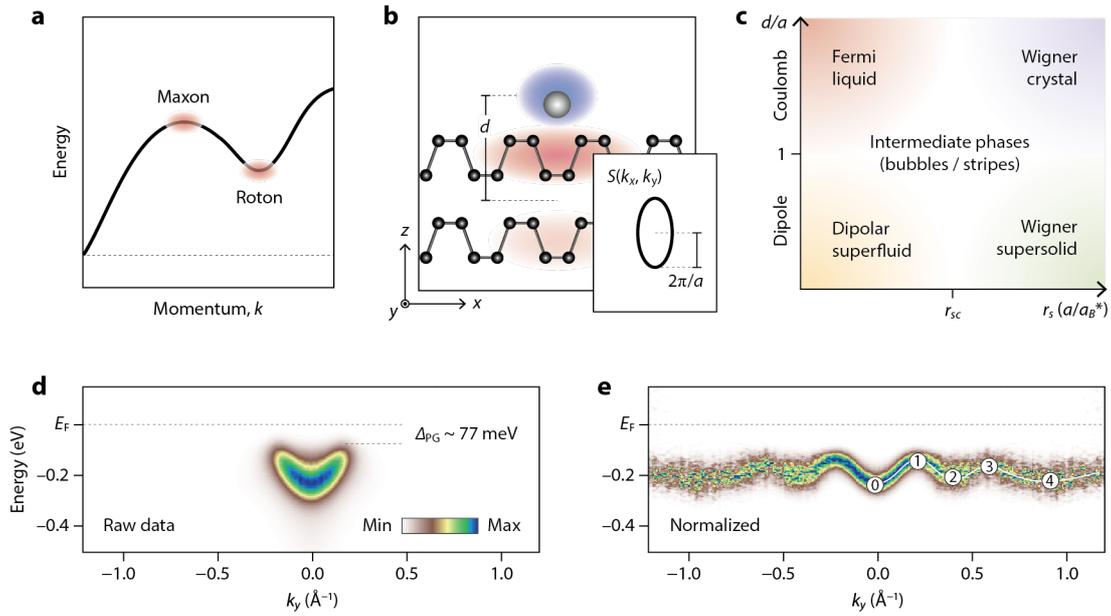

**Figure 1 | Electronic rotons in a 2D dipole liquid. a,** Landau's energy-momentum dispersion curve proposed for the excitation spectrum of liquid helium 4[1,2]. The dotted line indicates zero energy, and the light red ellipses represent the local maximum (maxon) and the local minimum (roton). **b,** Schematic illustration for our experimental system consisting of alkali metals (grey ball) on black phosphorus (black balls and sticks). Each alkali metal donates an electron to surface layers of black phosphorus, which forms a 2D dipole with the vertical distance of $d$. Inset shows $S(k_x, k_y)$ obtained from scanning tunnelling microscopy images[34] by Fourier transformation, providing information on the spatial arrangement of 2D dipoles. **c,** Schematic phase diagram for the 2D dipole system with two dimensionless parameters, $d/a$ and $r_s$ $(a/a_B^*)$ predicted in theory[22-27]. **d,** ARPES data of black phosphorus whose surface is doped by K at $n = 3.8 \times 10^{13}$ cm$^{-2}$, taken with the photon energy of 104 eV and the sample temperature of 25 K. The dotted lines indicate the pseudogap $\Delta_{PG} = 77$ meV. **e,** ARPES data obtained from **d** by normalising the maximum intensity of EDCs at each $k_y$ point. Each local maximum and minimum in the band dispersion is marked by the number in sequence.

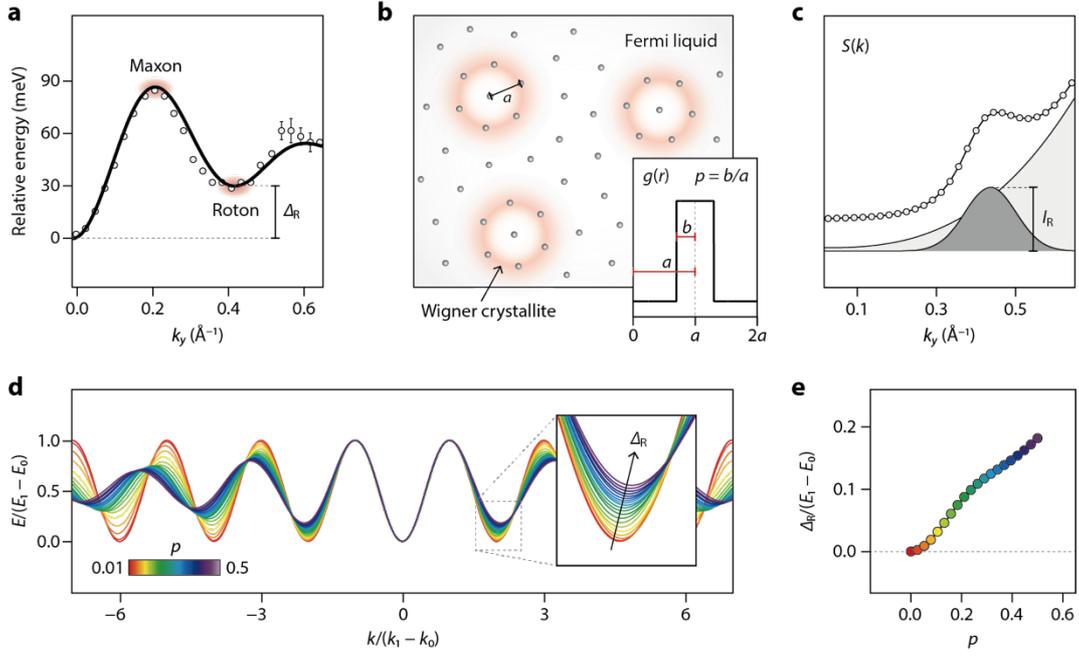

**Figure 2 | Analytic model based on the short-range order of dipoles. a,** Peak position (open circles) obtained by curve fitting to each EDC of ARPES data in Fig. 1d,e. Error bars show the maximum deviation of curve fittings with the population size of 300. The overlaid black curve is $E(k)$ calculated with the analytic model[30,31] (Methods). The local maximum (maxon) and the local minimum (roton) are marked in red, and dotted lines show the roton gap $\Delta_R$. **b,** Schematic illustration for Wigner crystallites floating in the sea of Fermi liquid: Grey balls represent dipoles, and some of them enclosed by red rings show a local microcrystal with the interatomic distance of $a$. Inset shows $g(r)$ considered in our model with a rectangular peak located at $a$ and whose width is $b$ ($p = b/a$). **c,** Experimental $S(k)$ shown by open circles and taken from peak positions in **a** through Feynman's relation[3]. This curve is decomposed into the Gaussian peak (dark grey) whose height is $I_R$ and backgrounds (light grey). **d,** Series of $E(k)$'s calculated based on the analytic model[30,31] (Methods) by varying $p$ defined in **b**, inset. **e,** Plot of $\Delta_R$ in **d** as a function of $p$ shown in the same colour as in **d**.

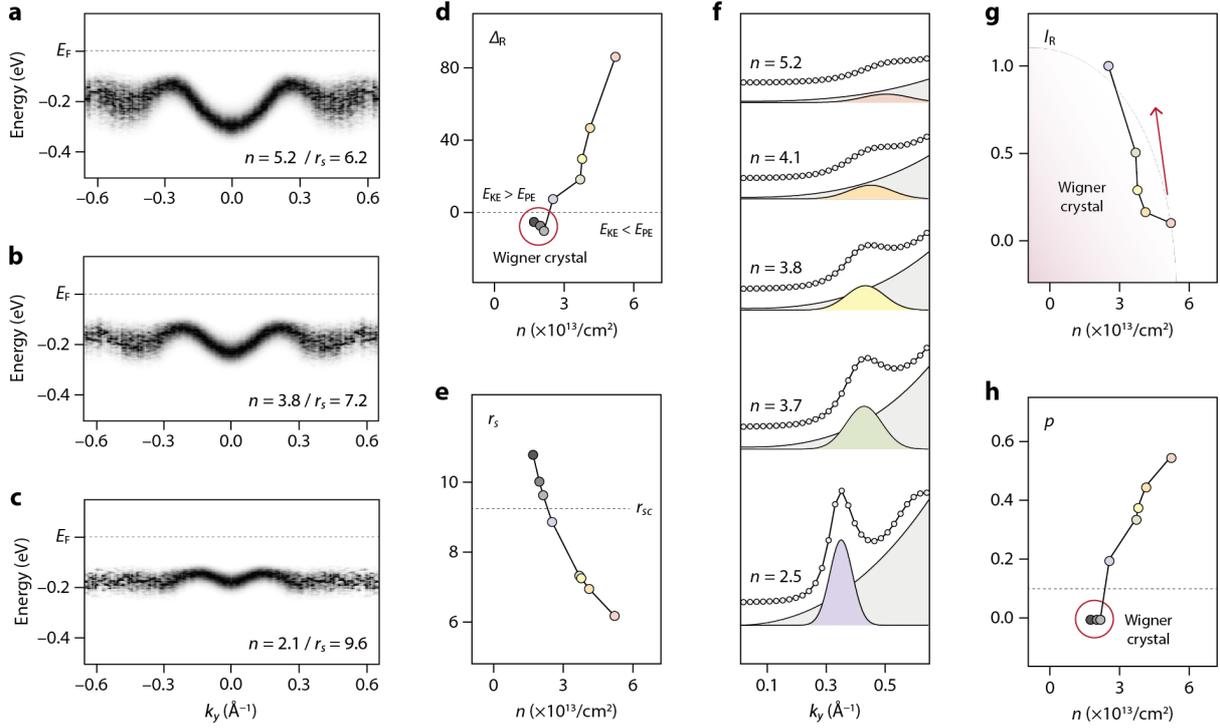

**Figure 3 | Wigner crystallisation. a-c,** Doping dependence of ARPES data taken from black phosphorus whose surface is doped by K at different $n$ marked at the bottom right in units of $10^{13}$ cm$^{-2}$ together with $r_s$ calculated by $a/a_B^*$. **d,e,** $\Delta_R$ (**d**) and $r_s$ (**e**) obtained by fitting to band dispersions in ARPES data (Extended Data Fig. 5) with $E(k)$ calculated by the analytic model[30,31] (Methods), and plotted as a function of $n$. The dotted line in **d** shows $n$ at which there is a crossover between $E_{KE} > E_{PE}$ and $E_{KE} < E_{PE}$. This $n$ is used to estimate $r_{sc}$ in **e** as shown by the same dotted line. **f,** Series of $S(k)$ taken through Feynman's relation[3] from ARPES data for $n$ marked at each set of data (open circles) in units of $10^{13}$ cm$^{-2}$. Each set of data is decomposed by curve fitting into the Gaussian peak in colour and the background in light grey (offset for clarity). **g,** $I_R$ of the Gaussian peak in **f** plotted as a function of $n$ and shown in the same colour as in **f**. The red arrow shows the development of the peak in $S(k)$. **h,** Plot of $p$ taken from $\Delta_R$ in **d** based on their relation in Fig. 2e. Red circles in **d,h** mark the onset of Wigner crystallisation.

## Methods

**Sample preparation.** We used a high-quality single crystal black phosphorus with the purity better than 99.995% (HQ graphene). The samples cut in the typical size of $3 \times 3 \times 0.5$ mm$^3$ were glued on sample holders with conductive silver epoxy. The samples prepared in this way were transferred into the ultrahigh vacuum chamber through the automated sample transfer system. The fresh surface of samples was prepared in situ by the standard cleaving method using the wobble stick at the ARPES chamber with the base pressure better than $5 \times 10^{-11}$ torr. The natural cleavage of black phosphorus is rather flaky, and we searched for a few spots, where we could find sharp ARPES spectra coming from the uniform domains using the well-focused synchrotron beam of 50 micrometre in diameter. We measured the reference spectrum of pristine black phosphorus (before surface doping)[51] for more than 500 samples with no exception to keep high consistency in the substrate. Before cleaving samples, we thoroughly outgassed evaporators used to deposit four kinds of alkali metals (Na, K, Rb, and Cs). These evaporators were equipped with alkali-metal dispensers (SAES), and one of the four kinds was deposited by electrothermal heating onto black phosphorus at 10–30 K from right below the inlet of the ARPES analyser. We used the fully automated deposition with a rectangular-shape voltage pulse to have high tuneability and consistency in the deposition rate. It is critical to optimise the quality of both samples and surface doing to the limit to resolve such a clear signature of electronic rotons as in Fig. 1e and Fig. 3a-c.

**ARPES experiment.** We carried out ARPES measurements at Beamline 7.0.2 (MAESTRO) of the Advanced Light Source (ALS). The microARPES end-station in MAESTRO is equipped with a hemispherical electron analyser (ScientaOmicron, R4000) and a home-made highly efficient deflector. The use of this deflector is found critical to finely tune the alignment of samples without losing the beam position. We used the *p*-polarised light of 104 eV, because it corresponds to a Z point of the bulk Brillouin zone, where the conduction band minimum of black phosphorus is located, but the choice of other photon energies and polarisations does not make difference in the band dispersion of electronic rotons. We collected ARPES spectra from our samples at the temperature of 10–30 K. With these experimental settings, the overall energy and angular resolutions were better than 15 meV and 0.1°, respectively. The 2D matrix data of ARPES intensity as a function of kinetic energies and emission angles was converted based on the energy and momentum conservations to that as a function of binding energies and *k*. The scale-converted data were symmetrised with respect to $k = 0$ and shown in Fig. 1d and Extended Data Figs. 2 and 3. Data in Fig. 1e, Fig. 3a-c, and Extended Data Figs. 2-5 are shown by normalising the maximum intensity of the EDCs at each *k* point. Just with this intensity normalisation, one can clearly see a band dispersion away from $\pm k_{PG}$ despite their relatively weak intensity because there exists a well-defined peak in each EDC with a moderate width and finite intensity (Extended Data Fig. 1). $E_F$ was calibrated based on the clear metallic bands that appear at the high density of alkali metals. The magnitude of $\Delta_{PG}$ in Fig. 1d and Extended Data Figs. 3 and 5 could be estimated by the energy distance from $E_F$ to the leading edge of the EDC at the Fermi wavevector ($k_F$) linearly extrapolated from the observed band dispersions.

**Structure factor.** The static structure factor $S(k)$ is experimentally determined from ARPES data based on the well-known Feynman relaltion[3]. We first fit each EDC of ARPES data with the standard Gaussian function to determine peak positions in energy as a function of $k$, as shown in Fig. 2a. The dispersion curve determined from ARPES data in this way was put in the equation, $E(k) = \hbar^2 k^2 / 2m^* S(k)$, to obtain $S(k)$'s shown in Figs. 2c and 3f. The first peak in $S(k)$ near $k = 2\pi/a$ is further analysed by curve fitting with the Gaussian function and the polynomial background. The height of this Gaussian peak defined by $I_R$ in Fig. 2c is plotted as a function of $n$ in Fig. 3g.

**Analytic model.** The aperiodic band dispersion is analysed based on theoretical models[30,31] for the electronic structure of non-crystalline systems in the presence of short-range order. The model is basically concerned with how a periodic band structure of crystalline systems changes by the effect of $g(r)$ that reflects the short-range order of noncrystalline systems. $g(r)$ is assumed as shown by inset in Fig. 2b to have rectangular-shape peaks only at $r = a$. The width of these peaks is $b$, and their heights are set to have the area of each peak unity (1/2b for the 1D case). Then, the ratio of $b$ to $a$ can be defined by a dimensionless parameter $p$ for the degree of short-range order or how uniform the average distance between atoms is. We first use the simplest linear chain of atoms with their atomic potential in the form of Dirac's $\delta$-function whose height is $\omega_0$. $E(k)$ was calculated using Equation (13) in Ref.[31] with input parameters of $a$, $p$ and $\omega_0$. A series of $E(k)$ obtained by varying $p$ is normalised based on maxon energy ($E_1 - E_0$) and maxon $k$ ($k_1 - k_0$) to show the $p$ dependence of $\Delta_R$ in Fig. 2d. For curve fitting to the band dispersion, since this model includes only the first-order peaks, we limit our attention to the first roton minimum as shown in Fig. 2a and Fig. 3a-c. We used a more sophisticated model with a spherical potential well whose width is $W$ and depth is $D$. $E(k)$ is calculated using Equation (78) in Ref.[31] with input parameters of $a$, $p$, $W$, and $D$. In this curve fitting, $a$ could be independently determined by maxon $k$ ($k_1 - k_0$) and roton $k$ ($k_2 - k_0$), and $p$ could be determined from the magnitude of $\Delta_R$ ($E_2 - E_0$). We set $D$ as the same value used in Ref.[33], which is 7.4 eV for Na, 16.4 eV for K and Rb, and 27.8 eV for Cs. Then, the only one remaining free parameter is $W$ that can be used to fit the overall energy scale of $E(k)$ as shown by the overlaid black curve in Fig. 2a.

**Dipole density and coupling parameter.** By fitting the band dispersion of ARPES data with $E(k)$ calculated by the model[30,31] described above, we could determine $a$ rather accurately. This $a$ could be used to quantify the density of dipoles $n$ in our system through the relation $n = 1/a^2 r_{xy}$, where $r_{xy}$ is the anisotropy ratio between $x$ (armchair) and $y$ (zigzag) directions. $r_{xy}$ is set to 3 from anisotropy in $k_F$ between $x$ and $y$ directions[33,34]. $n$ determined in this way has been used throughout this paper and in Figs. 1d,e and 3, and Extended Data Figs. 1–5. Furthermore, $a$ or $n$ could also be used to quantify the well-known dimensionless coupling parameter of 2D dipoles $r_s$ through the relation $r_s = a/a_B^*$ (Ref. [52]) or $\pi r_s^2 a_B^{*2} = 1/n$ (Ref. [22,23]), where $a_B^*$ is the effective Bohr radius. $a_B^*$ is taken by $4\pi\hbar^2\varepsilon/m^*e^2$, where $\varepsilon$ is the dielectric constant of monolayer black phosphorus comparable to the approximation for that of the interface between a vacuum and bulk black phosphorus. For the zigzag direction, we used $\varepsilon = 3.06\varepsilon_0$ from first principles calculations[53]. The effective mass $m^*$ is determined directly from ARPES data to be $1.25m_e$, which is in agreement with that predicted by first principle calculations[54]. $r_s$ determined in this way was used throughout this paper and in Fig. 3a-c,e and Extended Data Figs. 4 and 5. The reduction in $r_{sc}$ by the effect of defects or impurities was estimated based on theoretical calculations[55]. With the $d/a_B^*$ of 3–7 and the impurity density in the order of $10^{10}$ cm$^{-2}$ taken from STM data[56], $r_{sc}$ is estimated to be 6.7–11.8.

**ARPES simulation.** For the direct comparison between ARPES data and $E(k)$ calculated by theoretical models[30,31] we performed ARPES spectral simulations. The spectral function of ARPES intensity as a function of binding energy and $k$ is constructed based on the standard Lorentzian function. The binding energy of Lorentzian peaks at a given $k$ is obtained by $E(k)$ calculated based on the analytic model described above with parameters optimised to fit the band dispersion of corresponding ARPES data. The spectral width of Lorentzian peaks is set at 80 meV regardless of binding energy and $n$ for simplicity. The variation of intensity in raw data is simulated by the well-known destructive interference effect[57,58], which is the sigmoid function as shown by the red curve in Extended Data Fig. 1d. The simulated spectra in this way are directly compared with corresponding ARPES data in Extended Data Fig. 4.

**Method references**

51. Jung, S. W. *et al.* Black phosphorus as a bipolar pseudospin semiconductor. *Nat. Mater.* **19,** 277–281 (2020). https://doi.org/10.1038/s41563-019-0590-2
52. Golden, K. I., Kalman, G. J., Hartmann, P. & Donkó, Z. Dynamics of two-dimensional dipole systems. *Phys. Rev. E* **82,** 036402 (2010). https://doi.org/10.1103/PhysRevE.82.036402
53. Kutlu, E. Narin, P., Lisesivdin, S. B. & Ozbay, E. Electronic and optical properties of black phosphorus doped with Au, Sn and I atoms. *Phil. Mag.* **98,** 155–164 (2018). https://doi.org/10.1080/14786435.2017.1396375
54. Fei, R. & Yang, L. Strain-Engineering the Anisotropic Electrical Conductance of Few Layer Black Phosphorus. *Nano Lett.* **14,** 2884–2889 (2014). https://doi.org/10.1021/nl500935z
55. Chui, S. T. & Tanatar, B. Phase diagram of the two-dimensional quantum electron freezing with external impurities. *Phys. Rev. B* **55,** 9330–9932 (1997). https://doi.org/10.1103/PhysRevB.55.9330
56. Tian, Z. *et al.* Isotropic charge screening of anisotropic black phosphorus revealed by potassium adatoms. *Phys. Rev. B* **100,** 085440 (2019). https://doi.org/10.1103/PhysRevB.100.085440
57. Shirley, E. L., Terminello, L. J., Santoni, A. & Himpsel, F. J. Brillouin-zone-selection effects in graphite photoelectron angular distributions. *Phys. Rev. B* **51,** 13614–13622 (1995). https://doi.org/10.1103/PhysRevB.51.13614
58. Moser, S. An experimentalist's guide to the matrix element in angle resolved photoemission. *J. Electron Spectrosc. Relat. Phenom.* **214,** 29–52 (2017). https://doi.org/10.1016/j.elspec.2016.11.007

**Acknowledgements** This work was supported by National Research Foundation (NRF) of Korea funded by the Ministry of Science and ICT (Grants No. NRF-2021R1A3B1077156, NRF-RS-2024-00416976, and NRF-RS-2022-00143178) and the Yonsei Signature Research Cluster Program (2024-22-0163). This research also used resources of the Advanced Light Source, which is a DOE Office of the Science User Facility under Contract No. DE-AC02-05CH11231.

**Author contributions** S.P. performed ARPES experiments with help from C.J., E.R., and A.B, and carries out data analysis with help from M.H. K.S.K. conceived and directed the project. S.P. and K.S.K wrote the manuscript with contributions from all other co-authors.

**Competing interests** The authors declare no competing interests.

**Additional information**
**Correspondence and requests for materials** should be addressed to Keun Su Kim.
**Peer review information** *Nature* thanks --- for their contribution to the peer review of this work.
**Reprints and permissions information** is available at www.nature.com/reprints.

## Data availability

All relevant data are included with this paper as source data (Fig. 1d,e, Fig. 2a,c-e, Fig. 3a-h, Extended Data Fig. 1a-e, Extended Data Fig. 2b-g, Extended Data Fig. 3a-h, Extended Data Fig. 4a-f, and Extended Data Fig. 5a-f). Any other data that support findings of this paper are available from the corresponding author on request.

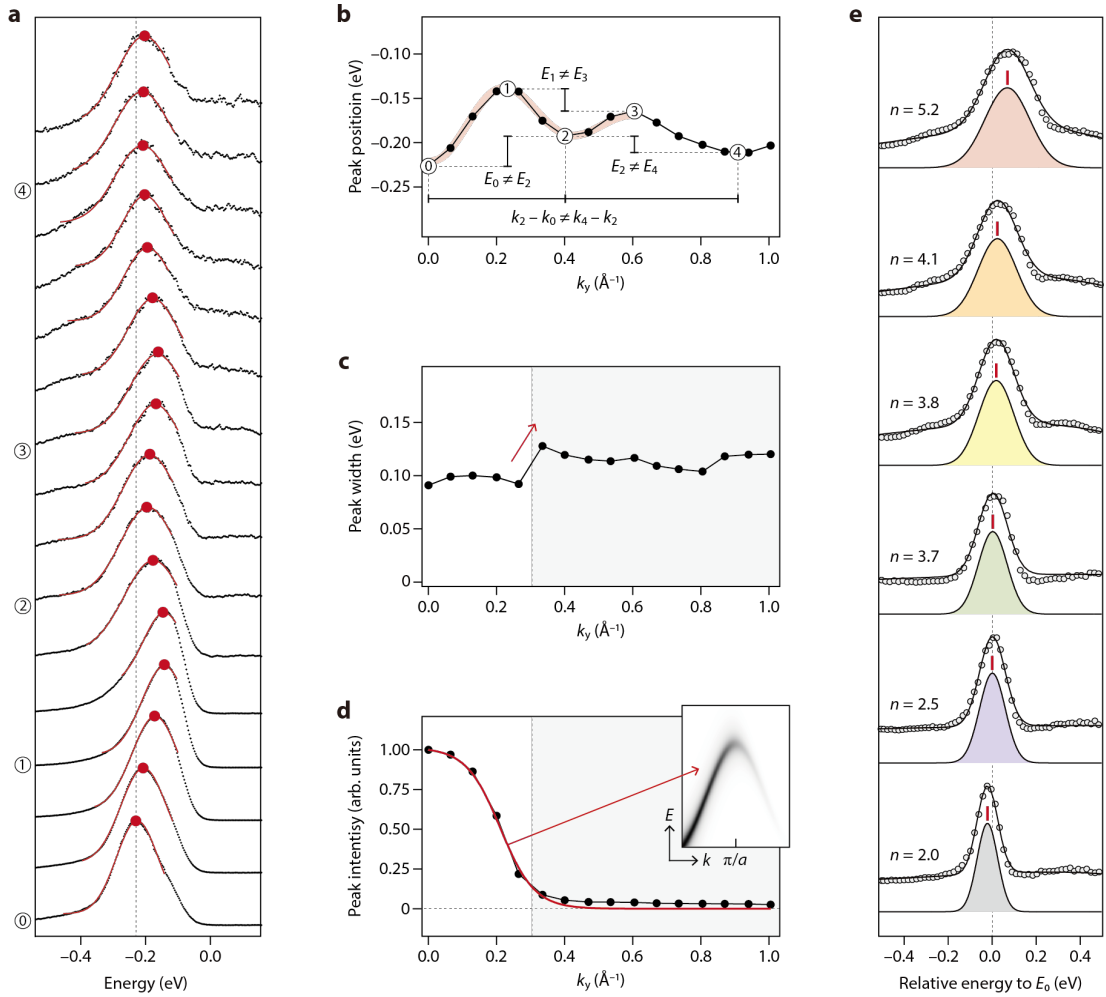

**Extended Data Fig. 1 | Raw ARPES data and curve fitting. a,** Raw data of Fig. 1d,e shown in the form of EDCs over the $k$ range from 0 to 1.005 Å$^{-1}$ with the $k$ interval of 0.067 Å$^{-1}$. There are well-defined peaks even in the $k$ region of weak intensity. For these well-defined peaks, it is straightforward by a fit to each EDC with the Gaussian function (red lines overlaid) to reliably determine their peak positions as marked by red circles relative to $E_0$ (dotted line). **b-d,** Peak position (**b**), width (**c**), and intensity (**d**) taken by curve fitting in **a** and plotted as a function of $k_y$. The error bars at maximum in **b** is smaller than the size of circled numbers. The light red line underlaid in **b** is a fit to peak positions with the band dispersion calculated by our model[30,31], which is in excellent agreement (quantitatively less than 4%). Even in the $k$ range of weak intensity shaded in grey, the peak width in **c** is only 20% greater than that of strong intensity (this is related to the imaginary part of $\Delta k$ in Extended Data Fig. 6a). On the other hand, the red curve overlaid in **d** shows the variation of intensity in conventional insulators across the zone boundary due to the well-known phase interference effect[57,58]. There is a small gap between peak intensity and the red curve, which means there remains relatively weak but finite intensity. **e,** Doping series of EDCs taken at the roton minimum or $k_2$ and plotted relative to $E_0$ (dotted line). Black lines overlaid are a fit to each EDC with the Gaussian peak in colour, where their peak positions are marked by red ticks.

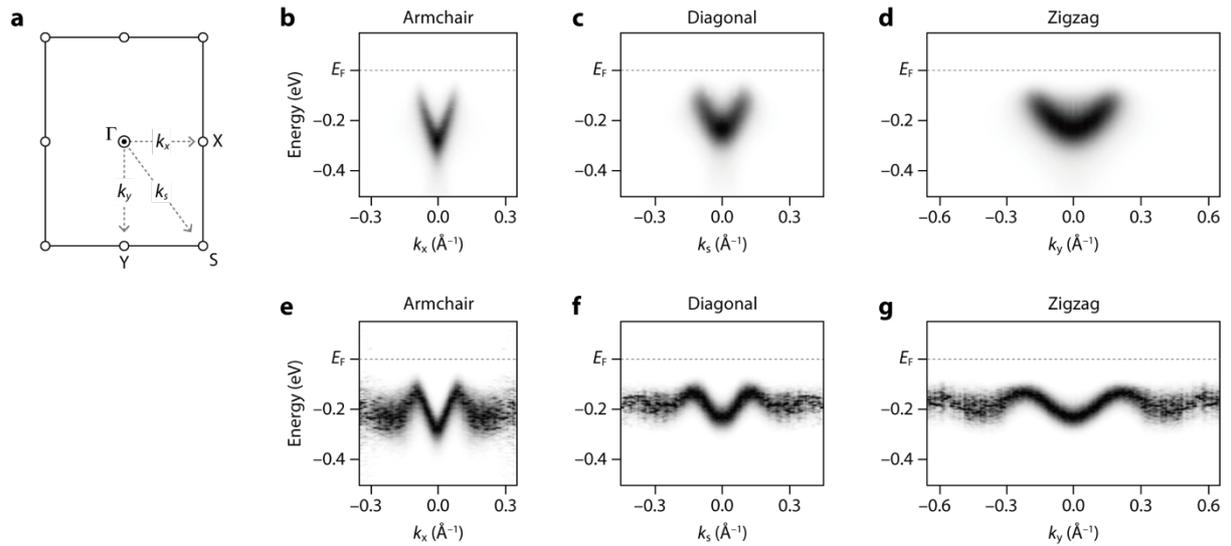

**Extended Data Fig. 2 | Direction dependence of electronic rotons. a,** Surface Brillouin zone of black phosphorus with high-symmetry points marked by open circles. Grey arrows show the three $k$ directions indexed by $k_x$, $k_s$, and $k_y$. **b-d,** ARPES data of black phosphorus doped by K at $n = 3.8 \times 10^{13}$ cm$^{-2}$, taken along armchair (**b**), diagonal (**c**), and zigzag (**d**) directions corresponding to $k_x$, $k_s$, and $k_y$, respectively. **e-g,** ARPES data obtained from those in **b-d** by normalising the maximum intensity of EDCs at each $k$ point. These normalised data reveal the clear signature of electronic rotons regardless of measurement directions.

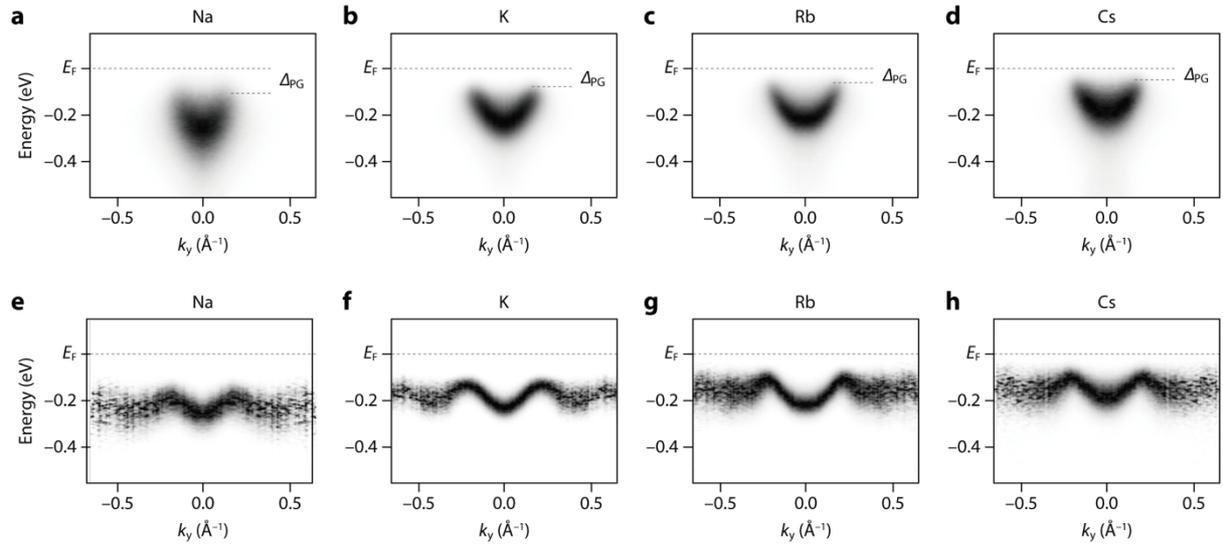

**Extended Data Fig. 3 | Element dependence of electronic rotons. a-d,** ARPES data taken in $k_y$ for black phosphorus doped by Na (**a**), K (**b**), Rb (**c**), and Cs (**d**) at $n = 3.8 \sim 5.3 \times 10^{13}$ cm$^{-2}$. Dotted lines show the magnitude of $\Delta_{PG}$, which is 106 meV for Na, 77 meV for K, 61 meV for Rb, and 48 meV for Cs as discussed in the previous report[33]. **e-h,** ARPES data obtained from those in **a-d** by normalising the maximum intensity of EDCs at each $k$ point. The normalised data show the clear signature of electronic rotons regardless of the kinds of alkali metals, which confirms that (*i*) it is not an artifact, and (*ii*) it is a generic property of alkali metals.

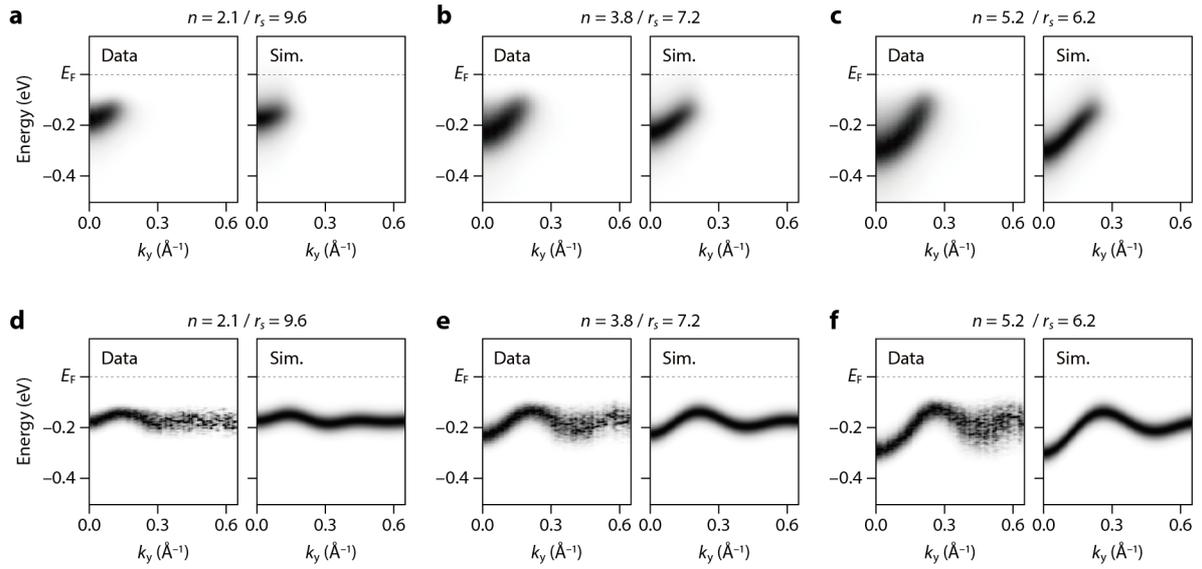

**Extended Data Fig. 4 | Comparison between ARPES data and simulations. a-c,** ARPES data (left panel) compared with simulations (right panel) for black phosphorus doped by K at $n$ marked on top of each panel in units of $10^{13}$ cm$^{-2}$ and $r_s$. The band dispersion for simulations is obtained by fitting the peak position of ARPES data and the variation of spectral intensity as a function of $k$ is obtained from the well-known phase interference effect[57,58] (Methods). **d-f,** Corresponding ARPES data and spectral simulations after normalising the intensity of peaks in each EDC. Our spectral simulations reproduce key aspects of not only raw data in **a-c,** but also normalised data in **d-f.**

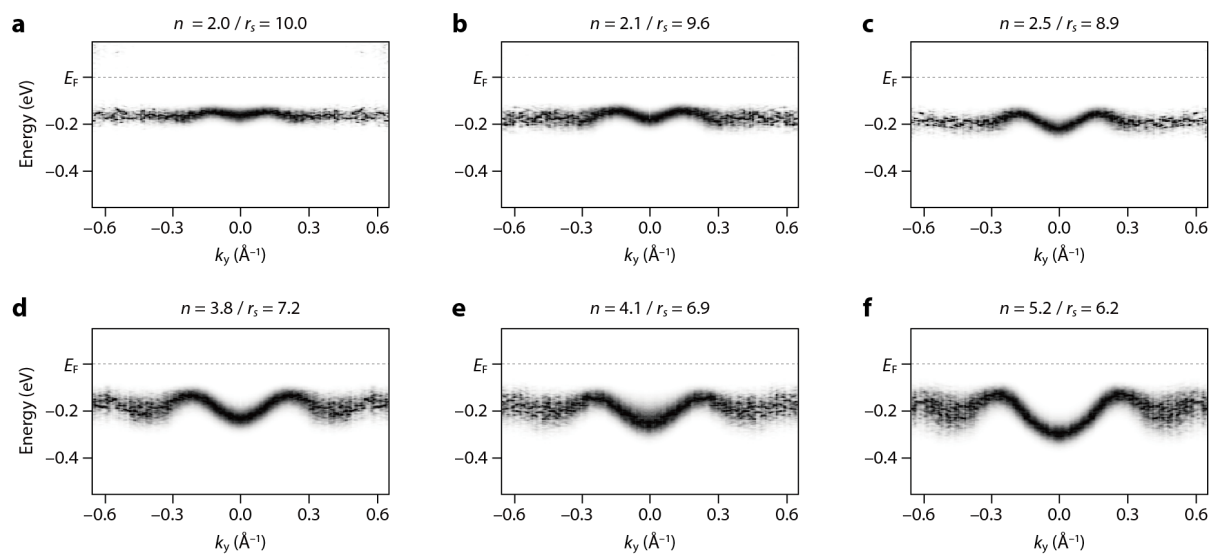

**Extended Data Fig. 5 | Doping dependence of electronic rotons. a-f,** ARPES data taken for black phosphorus doped by K at $n$ marked on top of each panel in units of $10^{13}$ cm$^{-2}$ with $r_s$. We found the pseudogap of 71–107 meV persistent in the range of $n$.

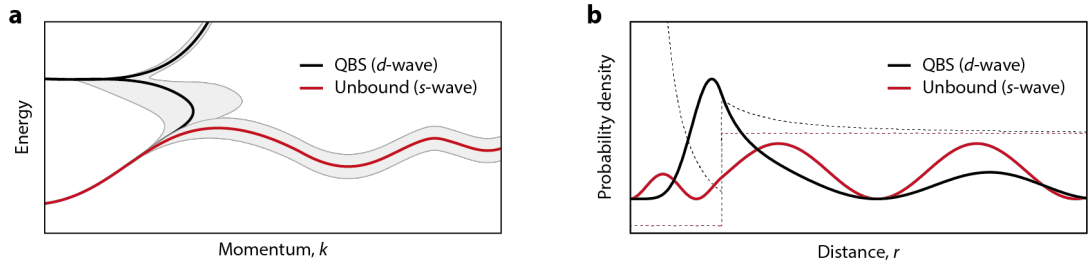

**Extended Data Fig. 6 | Complete electronic structure of liquid metals. a,b** Band dispersion (**a**) and probability density (**b**) of wavefunctions obtained by the theoretical model[46-49] that was initially developed for liquid metals but can be generally applied to any non-crystalline system in the presence of the short-range order. There are two branches in terms of partial wave analysis[46]: One is the *p*-wave or *d*-wave states at resonance scattering predicted by Anderson and McMillan[47] to show the back-bending band dispersion (due to the real part of $\Delta k$ as shown by the black curve) and the pseudogap (due to the imaginary part of $\Delta k$ as shown by the grey area), as shown in **a**. This is due to the formation of quasi-bound states (QBS), as shown by the black curve in **b**, within the scattering potential (dotted black line). The other is *s*-wave states[49], for which resonance scattering is forbidden by the absence of a potential barrier as shown by the red dotted line in **b**. The presence of the unbound states represented by the red curve in **b** was predicted by Schwartz and Ehrenreich[48] to be related to another aperiodic (damped oscillatory) branch in band dispersion that extends towards the zone boundary[30,31] as shown by the red curve in **a**. The grey region surrounding the red curve shows the imaginary part of $\Delta k$ that accounts for the small increase in the peak width of EDCs indicated by the red arrow in Extended Data Fig. 1c.